\documentclass[a4paper, 12pt, oneside]{article}
\usepackage[cp1251]{inputenc}
\usepackage[english, russian]{babel}
\usepackage{ graphics,amsfonts, amsmath,bm,amssymb, array}
\usepackage[dvips]{graphicx}
\usepackage[flushleft,small]{caption2}
\makeatletter

\renewcommand{\Re}{\mathop{\rm Re\,}}
\renewcommand{\Im}{\mathop{\rm Im\,}}

\makeatother {
\usepackage[usenames]{color}
\begin{document}
\thispagestyle{empty} \large
\renewcommand{\abstractname}{Abstract }
\renewcommand{\refname}{\begin{center}References\end{center}}
\renewcommand{\contentsname}{\begin{center}Contents\end{center}}
 \begin{center}
\bf
Exact solution of dispersion equation corresponding to ellipsoidal statistical
equation from Stokes' second  problem
\end{center}
\begin{center}
\bf
A. L. Bugrimov\footnote{$fakul-fm@mgou.ru$},
A. V. Latyshev\footnote{$avlatyshev@mail.ru$},
A. A. Yushkanov\footnote{$yushkanov@inbox.ru$}\\ and
V. A. Akimova\footnote{$vikont\_ava@mail.ru$}
\end{center}\medskip

\begin{center}
{\it Faculty of Physics and Mathematics,\\ Moscow State Regional
University, 105005,\\ Moscow, Radio str., 10--A}
\end{center}\medskip

\tableofcontents
\setcounter{secnumdepth}{3}
\renewcommand{\tablename}{\begin{center}Contents\end{center}}
\renewcommand{\contentsname}{\begin{center}Contents\end{center}}

\begin{abstract}
In the present work zero of dispersive function from Stokes' second problem
are investigated. Stokes' second problem is a problem about
behaviour of the rarefied gas filling half-space. A plane,
limiting half-space, makes harmonious oscillations
in the plane. The  linearization kinetic ellipsoidal the statistical equation
with parametre is used.
The factorization formula of dispersion function is proved.
By means of the factorization formula
zero of dispersion function in an explicit form and their research is carried out.
Dependence on dimensionless quantity collision frequency of a plane
limiting gas and on parametre equation  are investigated.

\end{abstract}

\begin{center}
\item{}\section*{Introduction}
\end{center}
\addcontentsline{toc}{section}{\bf Introduction}

Problem of generation of shift waves by fluctuating plate or Stokes' second
problem for the continuous matter  has been formulated in the middle
of XIX century \cite {Stokes}. Then, after introduction by Maxwell and
Boltzmann the kinetic equations, Stokes' second problem began to be studied
for the rarefied gas.

The detailed history of this problem is stated in our works
\cite {ALY-1}--\cite {ALY-3}. In these works the Stokes' second problem
was solved analytically. The known kinetic BGK--equation
(Bhatnagar, Gross, Krook) thus was  used.

In work \cite {Bugrimov} zero of the dispersion function corresponding to
BGK--equation have been investigated. The range of values of frequencies of
oscillations of the plane is found out, in which
to within one percent exact expression of zero
of dispersion function can replace it by asymptotic approximation
taken by by means of expansion in asymptotic series  of dispersion
function in neighbourhood of infinitely remote point.

Necessity of this replacement of exact expression of zero of dispersion function
by it asymptotic representation speaks that fact, that
calculation of macrocharacteristics of this problem demands variety calculation
of values of composite functions in zero of dispersion function.

Let's underline, that this problem draws to itself wide attention of many
authors (see, for example, \cite{Sharipov}--\cite{11}). It has been solved
by the numerical and (or) by approximate methods.
In works \cite{7,8} Stokes' second problem has been successfully applied in
nanotechnology.

In work \cite{ESv1} Stokes' second problem has been solved analytically with
help of ellipsoidal statistical equation.

Methods of calculation of zero of dispersion functions for the transport
equations of neutrons have been put in pawn in work \cite{Siewert0}.
Then these methods were applied and developed for various problems in
works \cite{Siewert1}--\cite{Anas4}.

In the present work in explicit form eigen solutions of ellipsoidal
statistical equation are presented. These solutions correspond to
discrete spectrum.
For this purpose in the explicit form zero of the dispersion equation
are found.
For finding of zero it is used factorization of dispersion function.
For this proved boundary value Riemann problem from the theory of functions
of complex variable. Coefficient of Riemann problem is
the relation of boundary values of dispersion function from above and from
below on the real semiaxis.

At small values of frequency of oscillations of a plane limiting rarefied gas
the simple asymptotic formula for calculation of zero of dispersion function
is found.
Graphic research of modules of zero of dispersion function is carried out,
and also the real and imaginary parts  calculated
by exact and asymptotic formulas.
The function of errors representing a relative deviation of the module
of asymptotic representations of zero from the module of its
exact representation is entered.

The interval of values of frequency of oscillations of a plane,
in which the value of function of errors does not exceed one percent
is found out.

In item 2 of the present work properties of dispersion function are studied in
complex plane. In item 3 exact formulas for calculation of zero
of dispersion function are deduced. Properties of these zero as functions
of dimensionless frequency of oscillations of the plane limiting rarefied
gas also are investigated.

In work \cite{ESv1} Stokes' second problem has been analytically solved with
using ellipsoidal statistical equation, which
after linearization and of some simplifications it is reduced to the equation
$$\textcolor{blue}{
\mu\dfrac{\partial h}{\partial x_1}+z_0h(x_1,\mu)=\dfrac{1}{\sqrt{\pi}}
\int\limits_{-\infty}^{\infty}\exp(-{\mu'}^2)(1-a\mu\mu')h(x_1,\mu')d\mu',}
\eqno{(1.1)}
$$
where
$$
z_0=1-i\Omega, $$
$x_1$ is the dimensionless coordinate, $x_1=x/l$, $l$ is the mean free path
of gaseous molecules, $\Omega=\omega\tau=\dfrac{\omega}{\nu}$,
$\tau=1/\nu$, $\nu$ is the frequency of collision of  gaseous molecules,
$\omega$ is the oscillation frequency of plates, limitting half-space
filling of rarefied gas, $a$ is the number parametre of problem, $0 \leqslant a
\leqslant 1$.

\begin{center}
\item{}\section{Statement problem}
\end{center}

Let rarefied one-atomic gas fills half-space $x>0$
over plane solid surface, laying in the plane $x=0$.
Surface $(y,z)$ makes harmonical oscillations lengthwise an axes $y$
under the law $u_s(t)=u_0e^{-i\omega t}$.

We will be linearize the distribution function of gaseous molecules
believing
$$
f(x,t,\mathbf{v})=f_0(v)(1+\varphi(x,t,\mathbf{v})).
$$
Here
$f_0(v)=n(\beta/\pi)^{3/2}\exp(-\beta v^2)$ is the absolute Maxwellian,
$\beta=m/(2kT)$,  $k$ is the Boltzmann constant, $T$ is the temperature of
gas,
$n$ is the concentration (number density) of gas, $m$ is the mass of
molecule of gas.

Let further $\nu=1/\tau$ is the collision frequency of gaseous molecules,
$\tau$ is the time between two consecutive collisions of molecules,
$u_y(x,t)$ is the mass velocity of gas, $\sigma_{xy}(x,t)$ is the
component of viscous stress tensor,
$$
u_y(x,t)=\dfrac{1}{n}\int v_yf(x,t,\mathbf{v})d^3v,
$$
$$
\sigma_{xy}(x,t)=m\int v_xv_y f(x,t,\mathbf{v})d^3v.
$$
Concentration of gas and its temperature are considered as constants
in linearization statement of problem.

We introduce dimensionless velocities and parametres:
dimensionless velocity of molecules
$\mathbf{C}=\sqrt{\beta}\mathbf{v}$ \;$(\beta=m/(2kT))$,
dimensionless velocity of gas $U_y(x,t)=\sqrt{\beta}u_y(x,t)$,
dimensionless time $t_1=\nu t$ and
dimension\-less velocity of surface $U_s(t)=U_0e^{-i\omega t}$,
dimensionless component of viscous stress tensor
$$
P_{xy}(x,t)=\dfrac{\beta}{\rho}\sigma_{xy}(x,t),
$$
where $U_0=\sqrt{\beta}u_0$ is the dimensionless amplitude of oscillation
velocity of half-space border. Then the linearization ellipsoidal statistical
kinetic equation (see, for example, \cite{Cerc}) (short: the ES--equation)
can be written down in the form
$$
\dfrac{\partial \varphi}{\partial t_1}+
C_x\dfrac{\partial \varphi}{\partial
x_1}+\varphi(x_1,t_1,\mathbf{C})=$$$$=
{2C_y}U_y(x_1,t_1)-2aC_xC_yP_{xy}(x_1,t_1).
\eqno{(1.1)}
$$

Here $a$ is the parametre of equation, and at $a=1$ Prandtl number
is true $(\Pr=2/3)$. Let's notice, that for dimensionless time
$U_s(t_1)=U_0e^{-i\Omega t_1}$.

Let's underline, that the problem about gas fluctuations resolves in
to linearization statement.
Linearization of problems it is spent on dimensionless mass speed
$U_y(x_1,t_1)$ provided that $|U_y (x, t_1)|\ll 1$. This inequality
is equivalent to the inequality
$$
|u_y(x_1,t_1)|\ll v_T,
$$
where $v_T=1/\sqrt{\beta}$ is the heat velocity of molecules,
having an order of velocity of a sound.

These quantities of dimensionless mass velocity and components of the viscous
stress tensor through function $\varphi$ are expressed as follows
$$
U_y(x_1,t_1)=\dfrac{1}{\pi^{3/2}}\int \exp(-C^2)C_y\varphi(x_1,t_1,
\mathbf{C})d^3C,
\eqno{(1.2)}
$$
and
$$
P_{xy}(x_1,t_1)=\dfrac{1}{\pi^{3/2}}\int \exp(-C^2)C_xC_y\varphi(x_1,t_1,
\mathbf{C})d^3C.
\eqno{(1.3)}
$$

Considering, that plate oscillations are considered along an axis $y $,
we will search function $\varphi $ in the form
$$
\varphi(x_1,t_1,\mathbf{C})=C_ye^{-i\Omega t_1}h(x_1,C_x).
\eqno{(1.4)}
$$
By means of (1.4) it is received the following boundary problem
$$\boxed{
\mu\dfrac{\partial h}{\partial x_1}+z_0h(x_1,\mu)
=\dfrac{1}{\sqrt{\pi}}
\int\limits_{-\infty}^{\infty}\exp(-{\mu'}^2)(1-a\mu\mu')h(x_1,\mu')d\mu',}
\eqno{(1.5)}
$$
$$
h(0,\mu)=2U_0,\qquad \mu>0, \qquad   z_0=1-i\Omega,
\eqno{(1.6)}
$$
$$
h(+\infty,\mu)=0.
\eqno{(1.7)}
$$
It is easy to show, that equation parametre $a$ and Prandtl number
are connected by equality
$$
\Pr=\dfrac{2}{2+a},\quad \text{from which}\quad a=\dfrac{2(1-\Pr)}{\Pr}.
$$

To the correct (true) Prandtl number  $ \Pr=2/3$ is answered value of parametre
$a=1$. At $a=0$ the ellipsoidal statistical equation passes in to
BGK--equation with Prandtl number $\Pr=1$, i.e. at $\Pr=1 \; a=0$.

\begin{center}
\item{}\section{Eigen solutions of continuous spectrum}
\end{center}

Separation of variables in the equation (1.5) is carried out
to the following substitution
$$
h_\eta(x_1,\mu)=\exp\Big(-\dfrac{x_1z_0}{\eta}\Big)\Phi(\eta,\mu),
\eqno{(2.1)}
$$
where $\eta$ is the parametre of separation, or spectral parametre, general
speaking, it is complex one.

Substituting (2.1) in the equation (1.5) it is received the characteristic
equation
$$
z_0(\eta-\mu)\Phi(\eta,\mu)=\dfrac{1}{\sqrt{\pi}}\eta n_0(\eta)-
\dfrac{1}{\sqrt{\pi}}a\mu\eta n_1(\eta),
\eqno{(2.2)}
$$
where
$$
n_k(\eta)=\int\limits_{-\infty}^{\infty}
\exp(-{\mu'}^2)\mu'^k\Phi(\eta,\mu')d\mu',\quad k=0,1.
$$

From the equation (2.2) we find, that it is possible to present it in
the form
$$
(\eta-\mu)\Phi(\eta,\mu)=\dfrac{1}{\sqrt{\pi}z_0}\eta n_0(\eta)(1-b \mu\eta),
\eqno{(2.3)}
$$
where
$$
b=-\dfrac{i\Omega a}{z_0}.
$$

Further we will accept the following normalizing condition
$$
n_0(\eta)\equiv \int\limits_{-\infty}^{\infty}
\exp(-{\mu'}^2)\Phi(\eta,\mu')d\mu'\equiv {z_0}.
$$
Then the equation (2.3) has at
$\eta,\mu\in(-\infty,+\infty)$ the following solution  \cite{Zharinov}
$$\boxed{
\Phi(\eta,\mu)=\dfrac{1}{\sqrt{\pi}}P\dfrac{\eta(1-b\mu\eta)}{\eta-\mu}+
e^{\eta^2}\lambda(\eta)\delta(\eta-\mu)},
\eqno{(2.4)}
$$
where $\delta(x)$ is the Dirac delta function, the symbol $Px^{-1}$
means an integral principal value at integration $x^{-1}$,
$\lambda(z)$ is the dispersion function, introducing by equality
$$\boxed{
\lambda(z)=1-i\Omega+\dfrac{z}{\sqrt{\pi}}\int\limits_{-\infty}^{\infty}
\dfrac{e^{-\tau^2}(1-b z\tau)d\tau}{\tau-z}}.
$$
This function can be transformed to the form
$$\textcolor{blue}{
\lambda(z)=-i\Omega+(1-b z^2)\lambda_0(z)},
$$
where $\lambda_0(z)$ is the known function from theory of plasma,
$$\textcolor{blue}{
\lambda_0(z)=\dfrac{1}{\sqrt{\pi}}\int\limits_{-\infty}^{\infty}
\dfrac{e^{-\tau^2}\tau d\tau}{\tau-z}}.
$$

Eigen functions (2.4) are called as eigen functions of continuous
spectrum for the spectral parametre $\eta$ fills continuously
all real axis.

Thus, eigen solution of the equation (1.5) look like (2.1),
in which function $\Phi(\eta,\mu) $ is defined by equality (2.4).

On the condition of our problem we search the solution which is not
increasing far from the wall.
In this connection the spectrum of the boundary problem we will name positive
real half-axes of parametre $\eta$.

Let's result Sokhotsky formulas from above and from below on the real axis
for dispersion function
$$
\lambda^{\pm}(\mu)=\pm i\sqrt{\pi}\mu e^{-\mu^2}(1-b\mu^2)-i\Omega+
\dfrac{1-b\mu^2}{\sqrt{\pi}}\int\limits_{-\infty}^{\infty}
\dfrac{e^{-\tau^2}\tau d\tau}{\tau-\mu}.
$$

\begin{figure}[h]
\begin{center}
\includegraphics[width=17.0cm, height=10cm]{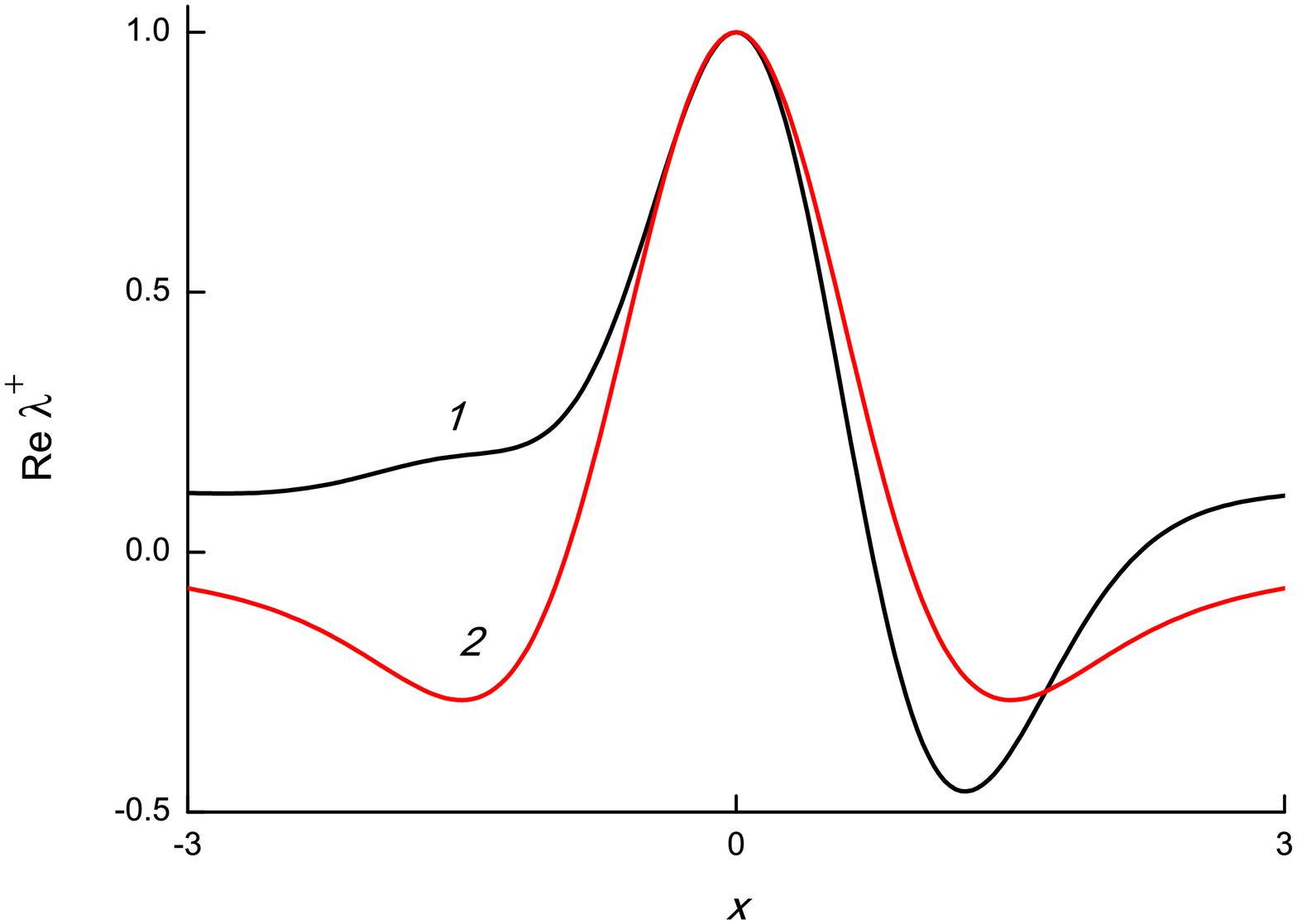}
\end{center}
\begin{center}
{{ Fig. 1. Real part of boundary values of dispersion function $\lambda^+(x)$,
frequency is equal to $\Omega=0.637$; curve 1 corresponds to
Prandtl number $\Pr=2/3$;  curve 2 corresponds to BGK--equation, $\Pr=1$.}}
\end{center}
\end{figure}

\begin{figure}[h]
\begin{center}
\includegraphics[width=17.0cm, height=9cm]{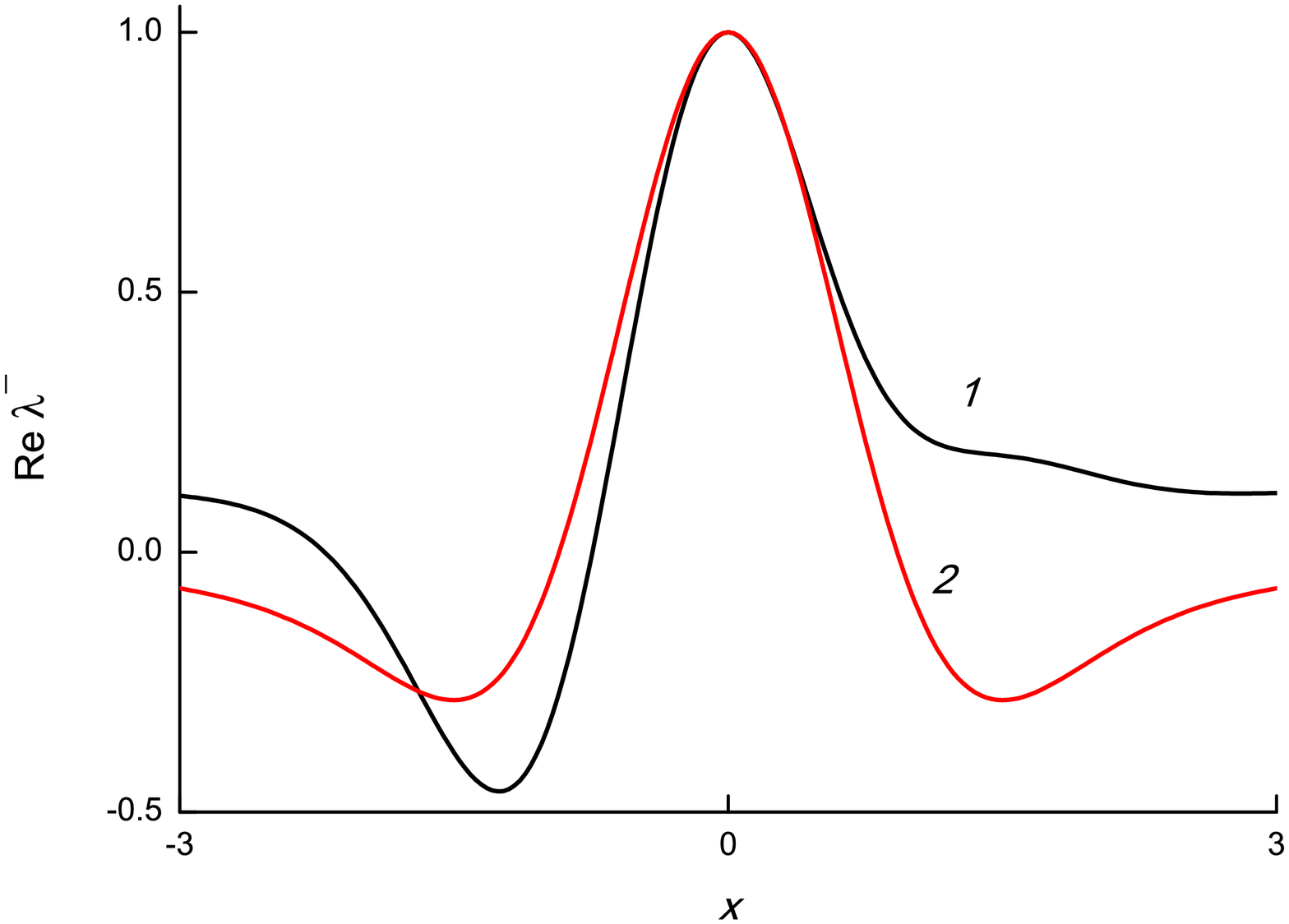}
\end{center}
\begin{center}
{{ Fig. 2. Real part of boundary values of dispersion function $\lambda^-(x)$,
frequency is equal to $\Omega=0.637$; curve 1 corresponds to
Prandtl number $\Pr=2/3$;  curve 2 corresponds to BGK--equation, $\Pr=1$.}}
\end{center}
\begin{center}
\includegraphics[width=17.0cm, height=9cm]{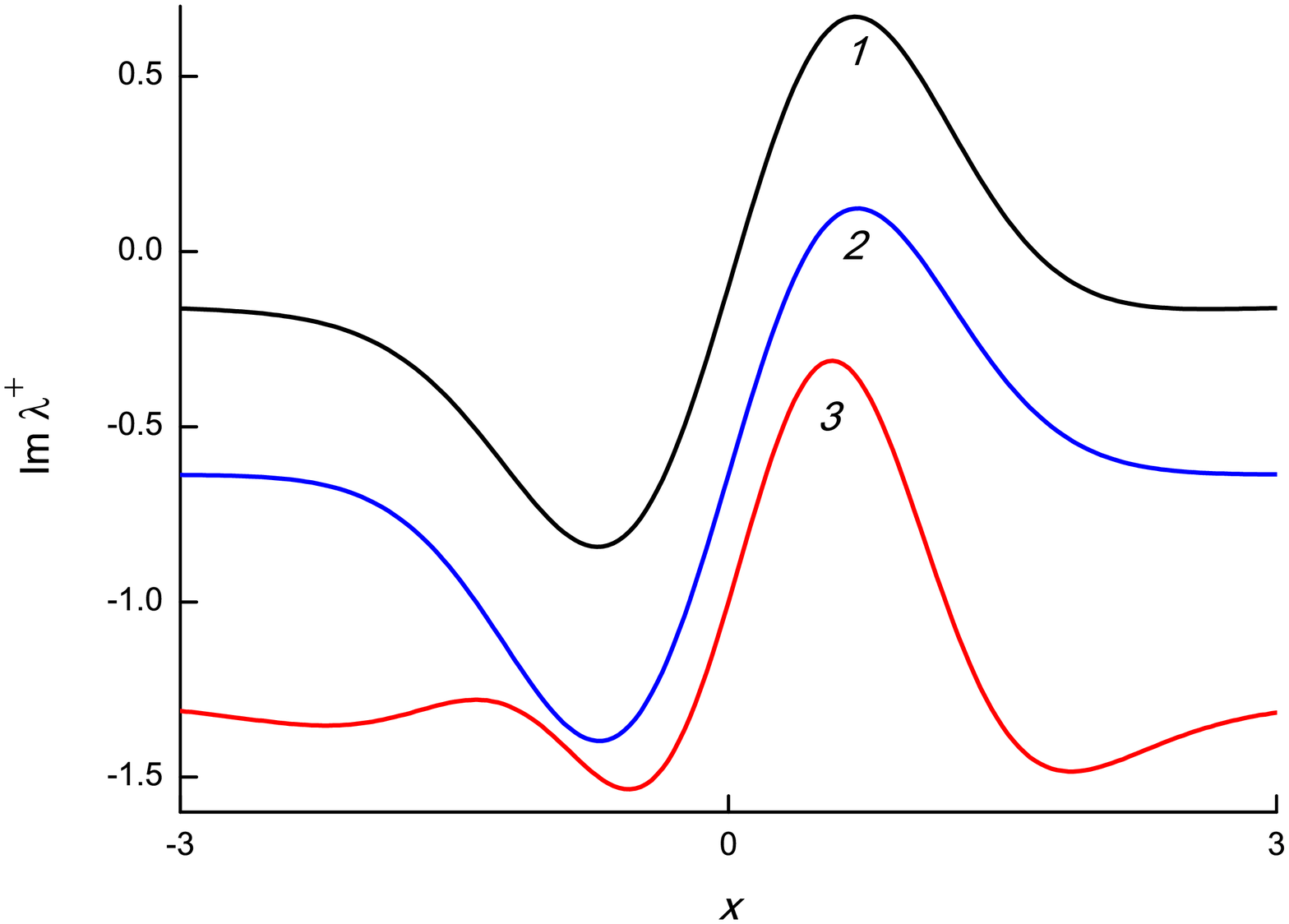}
\end{center}
\begin{center}
{{ Fig. 3. Imaginare part of boundary values of dispersion function
$\lambda^+(x)$; curve 1 corresponds to Prandtl number $\Pr=2/3$, $\Omega=0.1$;
curve 2 corresponds to $\Pr=2/3, \Omega=1$; curve 3 corresponds to
BGK--equation, $\Pr=1$, $\Omega=0.637$.}}
\end{center}
\end{figure}
\clearpage
\begin{figure}[h]
\begin{center}
\includegraphics[width=17.0cm, height=9cm]{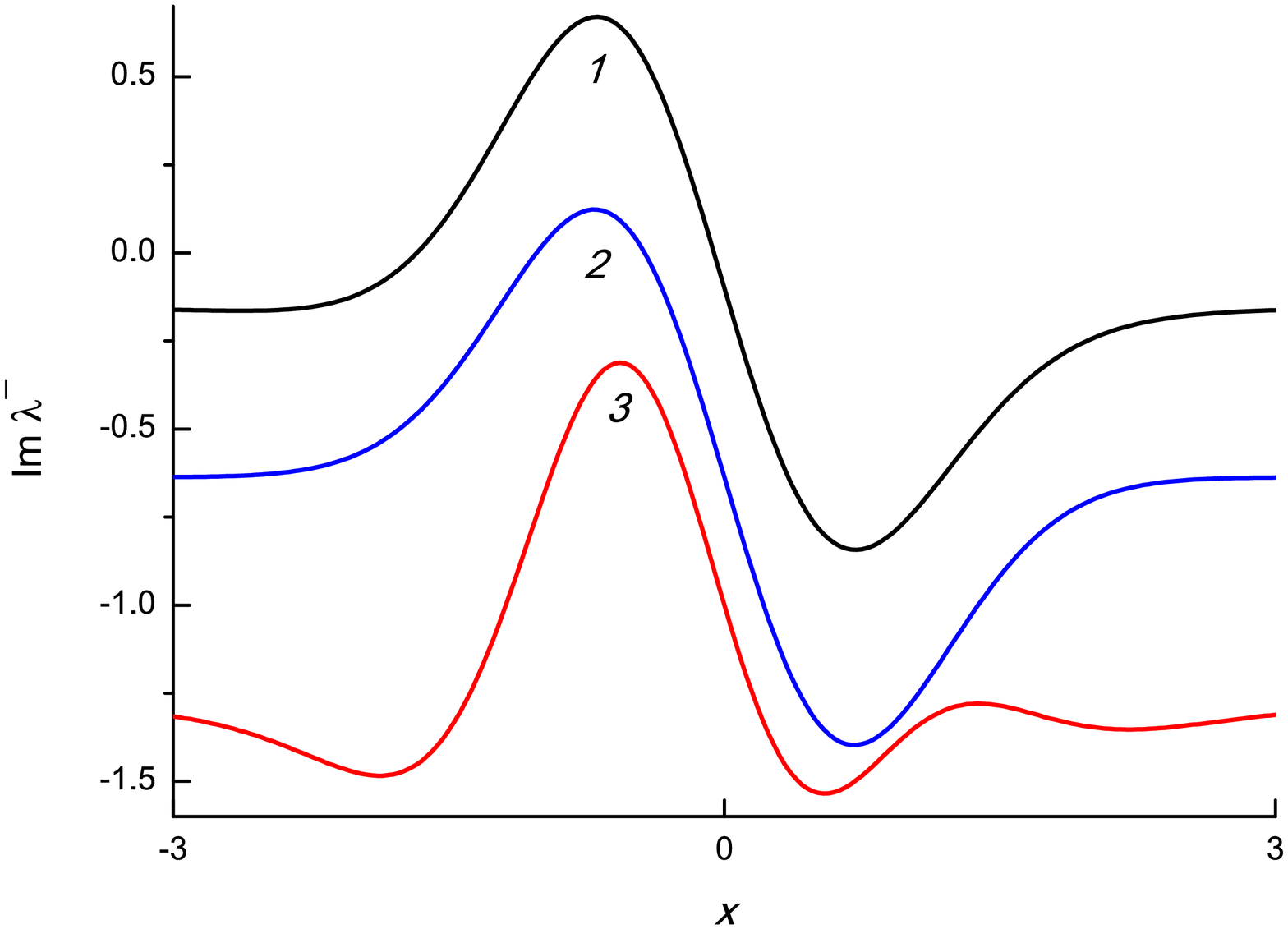}
\end{center}
\begin{center}
{{ Fig. 4. Imaginare part of boundary values of dispersion function
$\lambda^-(x)$, curve 1 corresponds to  Prandtl number $\Pr=2/3$, $\Omega=0.1$;
curve 2 corresponds to $\Pr=2/3, \Omega=1$; curve 3 corresponds to
BGK--equation, $\Pr=1$, $\Omega=0.637$.}}
\end{center}
\end{figure}

Difference of boundary values from above and from below on the real axis
to dispersion function from here it is equal
$$
\lambda^+(\mu)-\lambda^-(\mu)=2\sqrt{\pi}\mu e^{-\mu^2}(1-b\mu^2)i,
$$
the half-sum of boundary values is equal
$$
\dfrac{\lambda^+(\mu)+\lambda^-(\mu)}{2}=-i\Omega+\dfrac{1-b\mu^2}{\sqrt{\pi}}
\int\limits_{-\infty}^{\infty}\dfrac{e^{-\tau^2}\tau d\tau}{\tau-\mu}.
$$
The singular integral in these equalities is understood in sense of the
principal value.

\begin{center}
\item{}\section{Structure of  discrete spectrum}
\end{center}

Let's show, that the discrete spectrum consisting of zero of the
dispersion equations $\lambda(z)=0$, contains two zero $-\eta_0$ and $ \eta_0$,
from which are designated through $\eta_0$ that zero, at which $\Re \eta_0>0$.

At first we will consider the case of small values $ \Omega$.
Let's expand  dispersion function in asymptotic series on
negative degrees variable $z$ at the vicinity of infinitely remote point
$$
\lambda(z)=-i\Omega+\dfrac{b}{2}-\dfrac{1}{2z^2}+\dfrac{3b}{4z^4}+
\dots, \quad z\to \infty.
\eqno{(3.1)}
$$

From expansion (3.1) it is visible, that at small values $\Omega$
dispersion function has two complex zero differing only signs.
We will replace a number (3.1) its partial sum
$$
\lambda^{as}(z)=-i\Omega+\dfrac{b}{2}-\dfrac{1}{2z^2}.
$$
Then from the equation $\lambda^{as}(z)=0$ we will find asymptotic
representation of zero of the dispersion equation
$$\textcolor{Red}{
\pm\eta_0^{as}(\Omega)=
\sqrt{i\dfrac{1+3i\Omega a/2z_0}{\Omega(2+a/z_0)}}=
\sqrt{i\dfrac{1-i\Omega+3i\Omega a/2}
{2\Omega(1-i\Omega+a/2)}},\quad 0\leqslant a\leqslant 1.}
$$

From here it is visible, that at $ \Omega\to 0$ both zero of
dispersion function have as limit one infinitely remote point
$\eta_i =\infty $ of two order.

Now we investigate the case of any values $ \Omega$.
Further is required to us the function
$$
G(\tau)=\dfrac{\lambda^+(\tau)}{\lambda^-(\tau)}=
\dfrac{-i\Omega+(1-b\tau^2)\lambda_0^+(\tau)}
{-i\Omega+(1-b\tau^2)\lambda_0^-(\tau)}.
\eqno{(3.2)}
$$

Let's separate at function $G(\tau)$ the real and imaginary parts.
Let's notice, that
$$
b=b_1+ib_2,\qquad b_1=\dfrac{a\Omega^2}{1+\Omega^2},
\qquad b_2=-\dfrac{a\Omega}{1+\Omega^2},
$$
$$
\lambda_0^{\pm}(\tau)=l(\tau)\pm is(\tau),\quad
s(\tau)= \sqrt{\pi}\tau e^{-\tau^2},
$$
$$
l(\tau)=1-2\tau^2 \int\limits_{0}^{1}e^{-\tau^2(1-x^2)}dx.
$$

Now equality (3.2) can be written follows
$$
G(\tau)=
\dfrac{-i\Omega+[(1-b_1\tau^2)-ib_2\tau^2](l(\tau)+is(\tau))}
{-i\Omega+[(1-b_1\tau^2)-ib_2\tau^2](l(\tau)-is(\tau))},
$$
or
$$
G(\tau)=\dfrac{p+q-i(\Omega-p_1+q_1)}{p-q-i(\Omega+p_1+q_1)},
$$
where
$$
p(\tau)=(1-b_1\tau^2)l(\tau), \qquad
q(\tau)=b_2\tau^2s(\tau),
$$
$$
p_1(\tau)=(1-b_1\tau^2)s(\tau),\qquad
q_1(\tau)=b_2\tau^2l(\tau).
$$

Now the function $G(\tau)$ it is possible to present in the form
$$
G(\tau)=G_1(\tau)+iG_2(\tau),
$$
where
$$
G_1(\tau)=\dfrac{g_1(\tau)}{g_0(\tau)},\qquad
G_2(\tau)=\dfrac{g_2(\tau)}{g_0(\tau)},
$$
$$
g_1(\tau)=p^2-q^2+\Omega^2-p_1^2+q_1^2,
$$
$$
g_2(\tau)=2[pp_1+q(\Omega+q_1)],
$$
$$
g_0(\tau)=(p-q)^2+(\Omega+p_1+q_1)^2.
$$
These functions $g_j(\tau) (j=0,1,2) $ will be necessary in the explicit form
$$
g_1(\tau)=\Omega^2-[s^2(\tau)-l^2(\tau)][(1-b_1\tau^2)^2+b_2^2\tau^4],
$$
$$
g_2(\tau)=2s(\tau)\{\Omega b_2\tau^2+l(\tau)[(1-b_1\tau^2)^2+b_2^2\tau^4]\},
$$
$$
g_0(\tau)=\Omega^2+2\Omega[(1-b_1\tau^2)s(\tau)+b_2\tau^2l(\tau)]+
[l^2(\tau)+s^2(\tau)][(1-b_1\tau^2)^2+b_2^2\tau^4].
$$
In these equalities
$$
(1-b_1\tau^2)^2+b_2^2\tau^4=\dfrac{1+\Omega^2(1-a\tau^2)^2}{1+\Omega^2}.
$$
Thus, it is definitively received
$$
g_1(\tau)-\dfrac{\Omega^4-\Omega^2s_1(\tau)-s_0(\tau)}{1+\Omega^2},
$$
where
$$ s_0(\tau)=s^2(\tau)-l^2(\tau),\qquad s_1(\tau)=s_0(\tau)(1-a\tau^2)^2-1,
$$
and
$$
g_2(\tau)=\dfrac{2s(\tau)}{1+\Omega^2}\Big\{-a\Omega^2\tau^2+
l(\tau)[1+\Omega^2(1-a\tau^2)^2]\Big\},
$$
$$
g_0(\tau)=\Omega^2+[l^2(\tau)+s^2(\tau)]\dfrac{1+\Omega^2(1-a\tau^2)^2}
{1+\Omega^2}+
$$
$$
+2\Omega\Big[\Big(1-\dfrac{a\Omega\tau^2}{1+\Omega^2}\Big)
s(\tau)-\dfrac{a\Omega\tau^2}{1+\Omega^2}l(\tau)\Big].
$$

It is possible to show by means of principle of argument similarly that,
as it is made in \cite {ALY-1}, that number
of zero of dispersion function it is equal
$$
N=\dfrac{1}{2\pi i}\int\limits_{-\infty}^{\infty}d\ln G(\tau)=
\dfrac{1}{\pi i}\int\limits_{0}^{\infty}d\ln G(\tau)=$$$$=
\dfrac{1}{\pi}\Big[\arg G(\tau)\Big]_{0}^{+\infty}=\dfrac{1}{\pi}\arg G(+\infty)
=2\varkappa(G),
$$
i.e. to the doubled index of function $G (\tau)$.

Let's enter the angle $\theta(\tau)=\arg G(\tau)$, which is the principal
value of argument, fixed in zero by condition $ \theta(0) =0$,
$$
\theta(\tau)=\arcctg \dfrac{\Re G(\tau)}{\Im G(\tau)}=
\arcctg\dfrac{g_1(\tau)}{g_2(\tau)}.
\eqno{(3.3)}
$$
\begin{figure}[h]
\begin{center}
\includegraphics[width=17.0cm, height=9cm]{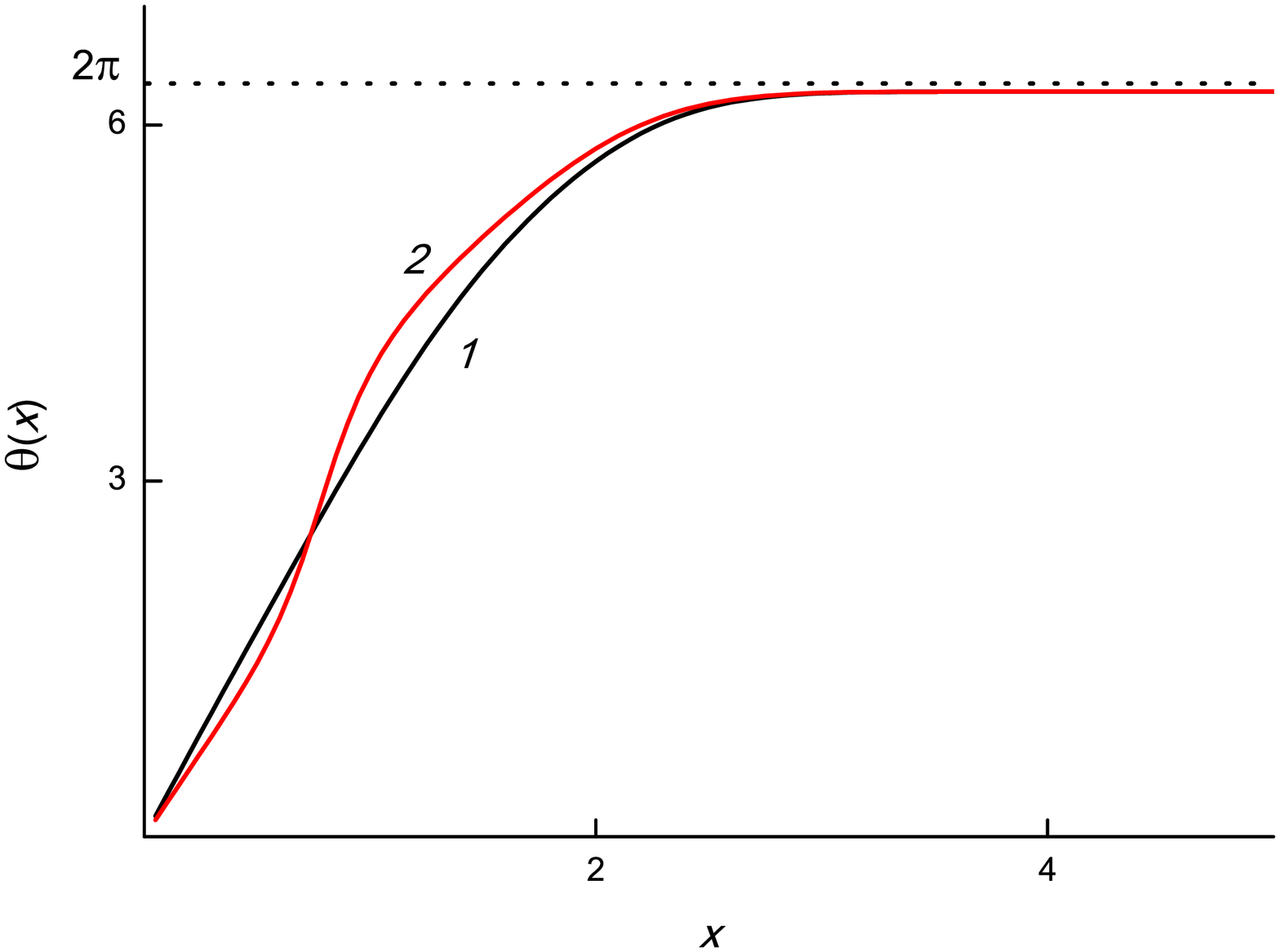}
\end{center}
\begin{center}
{{ Fig. 5. Angle $\theta=\theta(x)$, $\Pr=2/3$, curve 1 corresponds to
$\Omega=0.1$; curve 2 corresponds to $\Omega=0.5$.}}
\end{center}
\end{figure}

From equation $g_1(\tau)=0$ we find its positive root
$$\textcolor{blue}{
\Omega(a)=\sqrt{\dfrac{s_1(\tau)}{2}+
\sqrt{\Big(\dfrac{s_1(\tau)}{2}\Big)^2+
s_0(\tau)}}\equiv\Omega(\tau,a)}.
$$

Let's enter the allocated frequency of oscillations of the plate limiting gas
$$\boxed{
\Omega^*(a)=\max\limits_{0<\tau<\infty}\Omega(\tau,a)}.
\eqno{(3.4)}
$$
This frequency of oscillations we will name {\it critical}.

Similarly \cite{ALY-1} it is possible to show, that in the case, when frequency
plate oscillations less than critical, i.e. at $0\leqslant \omega <\Omega^*(a)$,
the index of function $G(t)$ is equal to unit.
It means, that number of complex zero of dispersion function in a plane
with cut along  real axis, equally to two.

In the case when frequency of oscillations of the plate exceeds the critical
($\omega>\Omega^*(a)$) the index of function $G(t)$ is equal to zero:
$ \varkappa(G)=0$.
It means, that dispersion function has no zero in top and bottom
of half-planes. In this case discrete (partial) solutions the initial
equation (1.9) has no.

Thus, the discrete spectrum of the characteristic equation,
consisting of zero of dispersive function, in the case
$0\leqslant \Omega <\Omega^*(a)$ there is a set from two points
$\eta_0$ and $-\eta_0$. At $\Omega>\Omega^*(a) $ the discrete
spectrum is the empty set. At $0 \leqslant \Omega <\Omega^*(a)$
decreasing eigen solution of the equation (1.9) looks like
$
h_{\eta_0}(x_1,\mu)=e^{-x_1z_0/\eta_0}\Phi(\eta_0,\mu),
$
where
$$
\Phi(\eta_0,\mu)=\dfrac{1}{\sqrt{\pi}}\dfrac{\eta_0(1-b\mu\eta_0)}{\eta_0-\mu}
$$
is the eigen solution of characteristic equation.

It means, that the discrete spectrum of the considered boundary problem
consists of one point $\eta_0$ in the case $0 <\Omega<\Omega^*(a)$.

At $ \Omega\to 0$ both zero $\pm\eta_0$ as already it was specified above,
move to one and same infinitely remote point. It means, that in the case
$\Omega=0$ the discrete spectrum of this problem consists of one infinitely
remote point of frequency rate two
also is attached to the continuous spectrum.
In this case discrete (partial) solutions exactly two:
$$
h_1(x_1,\mu)=1, \qquad h_2(x_1,\mu)=x_1-\dfrac{2}{2+a}\mu.
$$

Let's result the table of critical frequencies depending on values of
Prandtl number and equation parametre $a$ according to (3.4).

{\bf Table} of critical frequency values .\medskip
\bigskip

\begin{tabular}{|c|c|c|}
  \hline
Prandtl number $\Pr$ & Parametre $a$&Critical frequency $\Omega^*$\\\hline
  1 & 0 & 0.733 \\\hline
  0.952 & 0.1 & 0.717 \\\hline
  0.909 & 0.2 & 0.717 \\\hline
  0.870 & 0.3 & 0.691 \\\hline
  0,833 & 0.4 & 0.681 \\\hline
  0.800 & 0.5 & 0.672 \\\hline
  0.769 & 0.6 & 0.662 \\\hline
  0.741 & 0.7 & 0.654 \\\hline
  0.714 & 0.8 & 0.648 \\\hline
  0.690 & 0.9 & 0.642 \\\hline
  2/3   &   1 & 0.637 \\\hline
\end{tabular}

\begin{center}
\item{}\section{Factorization of dispersion function}
\end{center}

Here we deduce the formula representing factorization of
dispersion function in the top and bottom half-planes, and
also the formula for factorization boundary values  of dispersion function
from above and from below on the real axis is deduced.
Such factorization it is given in terms of function $X(z)$.

At the heart of the analytical solution of boundary problems
the kinetic theory  lays the solution of the homogeneous boundary
value Riemann problem (see \cite{Gakhov}) with
coefficient $G(\mu)=\lambda^+(\mu)/\lambda^-(\mu)$
$$
\dfrac{X^+(\mu)}{X^-(\mu)}=G(\mu), \; \qquad \mu> 0.
$$
Homogeneous boundary value Riemann problem is called also (see \cite{Gakhov})
{\it the factorization problem of coefficient} $G(\mu)$.

The Riemann problem mean that relationship $\lambda^+( \mu)/ \lambda^-( \mu)$
can rep\-la\-ce by relation $X^+( \mu)/X^-( \mu)$. Here $X^{\pm}(\mu)$
are boundary value of function $X(z)$, analytical in complex plane
$\mathbb{C}$ and having as jumps line the real positive half-axes.
Dispersion function have as line of jumps all real axis.

We consider  corresponding homogeneous boundary value Riemann problem
$$
X^+(\mu)=G(\mu)X^-(\mu),\qquad \mu>0,
\eqno{(4.1)}
$$
where coefficient of problem $G(\tau)$ is defined by equality (3.2).

The solution of Riemann problem Римана (4.1) is carried out similarly
\cite{ALY-2} and given by integral of Cauchy type
$$
X(z)=\dfrac{1}{z^\varkappa}\exp V(z),
\eqno{(4.2)}
$$
where $\varkappa=\varkappa(G)$ is the index of coefficient of problem,
entered in item 3, and $V(z)$ is understood as integral Cauchy type
$$
V(z)=\dfrac{1}{2\pi i}\int\limits_{0}^{\infty}
\dfrac{\ln G(\tau)-2\pi i \varkappa}{\tau-z}d\tau.
\eqno{(4.3)}
$$

Here $\ln G(\tau)=\ln |G(\tau)|+i \theta(\tau)$ is the principal branch of
logarithm, fixed at zero by condition $\ln G(0)=0$, angle
$\theta(\tau)=\arg G(\tau)$ is the principal value of argument,
entered by equality (3.3). The integral (4.3) is more convenient to consider in
the form
$$
V(z)=\dfrac{1}{\pi}\int\limits_{0}^{\infty}\dfrac{q(\tau)-\pi \varkappa}
{\tau-z},
$$
where
$$
q(\tau)=\dfrac{\theta(\tau)}{2}-\dfrac{i}{2}\ln|G(\tau)|,
$$
or in the form
$$
V(z)=\dfrac{1}{\pi}\int\limits_{0}^{\infty}\dfrac{\zeta(\tau)d\tau}
{\tau-z}, \qquad\varkappa=0,1,
$$
where
$$
\zeta(\tau)=q(\tau)-\pi \varkappa.
$$

Let first $\varkappa(G)=1$, i.e. $\Omega\in [0,\Omega^*(a))$.
We show that for dispersion function $ \lambda(z)$ everywhere in the
complex plane $ \mathbb{C}$, excepting the real  axis
$ \mathbb{R}$, is true the formula
$$\textcolor{blue}{\boxed{
\lambda(z)= -\lambda(\infty)(z^2-\eta_0^2)X(z)X(-z)}}.
\eqno{(4.4)}
$$

Here
$$
\lambda(\infty)=-i\Omega+\dfrac{b}{2}=
-i\Omega\Big[1+\dfrac{a}{2(1-i\Omega)}\Big]=
$$
$$
=-\dfrac{i\Omega}{\Pr}
\dfrac{1-i\Omega\Pr}{1-i\Omega},\qquad \dfrac{2}{3}\leqslant \Pr \leqslant 1.
$$

From this formula follows that for its boundary values on
$ \mathbb{R}$ are carried out the following relations
$$\textcolor{blue}{
\lambda^{\pm}( \mu)=-\lambda(\infty)(\mu^2-\eta_0^2)X^{\pm}( \mu)X(- \mu), \qquad
\mu \geqslant 0,}
\eqno{(4.5)}
$$
$$\textcolor{blue}{
\lambda^{\mp}( \mu)=-\lambda(\infty)(\mu^2-\eta_0^2)X( \mu)X^{\mp}(- \mu), \qquad
\mu \leqslant 0.}
\eqno{(4.6)}
$$

For the proof of the formula (4.4) we will enter auxiliary function
$$
R(z)= \dfrac{\lambda(z)}{-\lambda(\infty)(z^2-\eta_0^2)X(z)X(-z)}.
\eqno{(4.7)}
$$
This function analytic everywhere in the complex plane, except points of cuts
$ \mathbb{R_+}$ and $ \mathbb{R_-}$. These points $z=\pm \eta_0$ are
removable, because at these points $\lambda(\pm \eta_0)=0$.

Each point of cuts $ \mathbb{R_+}$ and $ \mathbb{R_-}$ is removable.
Really, if $ \mu>0$, then on the basis of equality (4.4) and (4.7) it is had
$$
\dfrac{ \lambda^+( \mu)}{\lambda(\infty)(\mu^2-\eta_0^2)X^+( \mu)X(- \mu)}=
\dfrac{ \lambda^-( \mu)}{\lambda(\infty)(\mu^2-\eta_0^2)X^-( \mu)X( -\mu)},
$$
from which $ R^+( \mu)= R^-( \mu), \quad \mu>0$.
If $ \mu<0$, then on the basis of equality (4.4), in which $ \mu$
is  replaced on $-\mu$, we have
$$
\dfrac{X^+( - \mu)}{X^-( - \mu)}=
\dfrac{ \lambda^+( \mu)}{ \lambda^-( \mu)},
\qquad \mu<0.
 $$
It easy to see that
 $
\lambda^+( - \mu)= \lambda^-( \mu), \quad \lambda^-( - \mu)=
\lambda^+( \mu).
$
Hence
$$
\dfrac{X^+( - \mu)}{X^-(- \mu)}= \dfrac{ \lambda^-( \mu)}
{ \lambda^+( \mu)}, \qquad \mu<0,
$$
from which
$$
\dfrac{\lambda^+( \mu)}{\lambda(\infty)(\mu^2-\eta_0^2)X( \mu)X^-(-\mu)}=
\dfrac{\lambda^-( \mu)}{\lambda(\infty)(\mu^2-\eta_0^2)X( \mu)X^+(-\mu)}, \qquad
\mu<0,
$$
or
$$
R^+( \mu)=R^-( \mu ), \qquad \mu<0.
$$

To prove equalities
$$
R^+( \mu)=
\dfrac{ \lambda^+( \mu)}{-\lambda(\infty)(\mu^2-\eta_0^2)X( \mu)X^-(-\mu)}
$$
and
$$
R^-( \mu)=
\dfrac{\lambda^-( \mu)}{-\lambda(\infty)(\mu^2-\eta_0^2)X(\mu)X^+(-\mu)},
$$
we notice that if poin $z$ tends to point $ \mu\;( \mu<0)$ from upper
or below half-plane, then functions $R^+( \mu)$
or $R^-( \mu)$ are calculated according to the previous equalities.

Hence, it is possible to consider this function as analytical function
everywhere in $ \mathbb {C} $, and in cut points, having predetermined it
on the cut by continuity. It is necessary to notice, that function $R (z)$
is analytical everywhere in $\mathbb{\overline{C}} $ and $R(\infty)=1$.
Under Liouville theorem  this function is
identically constant: $R (z)\equiv 1$,
whence the formula (4.4) is proved.

Formulas (4.5) and (4.6) obviously follow from the formula (4.4).

From the formula (4.4) we will find in the explicit form the formula
for calculation of zero of dispersion function
$$\boxed{
\eta_0(\Omega)=\sqrt{z^2+\dfrac{\lambda(z)}{\lambda(\infty)X(z)X(-z)}}}.
$$

In this formula as a point $z $ it is convenient to take the
point on the imaginary axis:
$z=Ni, N=1,2,\cdots$. Then we will receive the formula
$$
\eta_0(\Omega)=\sqrt{-N^2+\dfrac{\lambda(Ni)}{\lambda(\infty)X(Ni)X(-Ni)}}.
$$

We will calculate both parts of equality (4.4)
at the point $z=i $. As a result for zero of dispersion function it is received
the following formula
$$\boxed{
\eta_0(\Omega)=\sqrt{-1+\dfrac{\lambda(i)}{\lambda(\infty)}\exp\big[-V(i)-
V(-i)\big]}}.
\eqno{(4.8)}
$$

Let's consider the case of zero index: $\varkappa(G)=0$, i.e.
$\Omega\in (\Omega^*,+\infty)$.  Similarly previous are proved
formulas
$$\textcolor{blue}{
\lambda(z)=-\lambda(\infty)X(z)X(-z), \qquad \Im z \ne 0.}
$$
$$\textcolor{red}{
\lambda^{\pm}(\mu)=-\lambda(\infty)X^{\pm}(\mu)X(-\mu), \qquad \mu \leqslant 0.}
$$
$$\textcolor{magenta}{
\lambda^{\pm}(\mu)=-\lambda(\infty)X(\mu)X^{\mp}(-\mu), \qquad \mu <0.}
$$

\begin{figure}[h]
\begin{center}
\includegraphics[width=17.0cm, height=10cm]{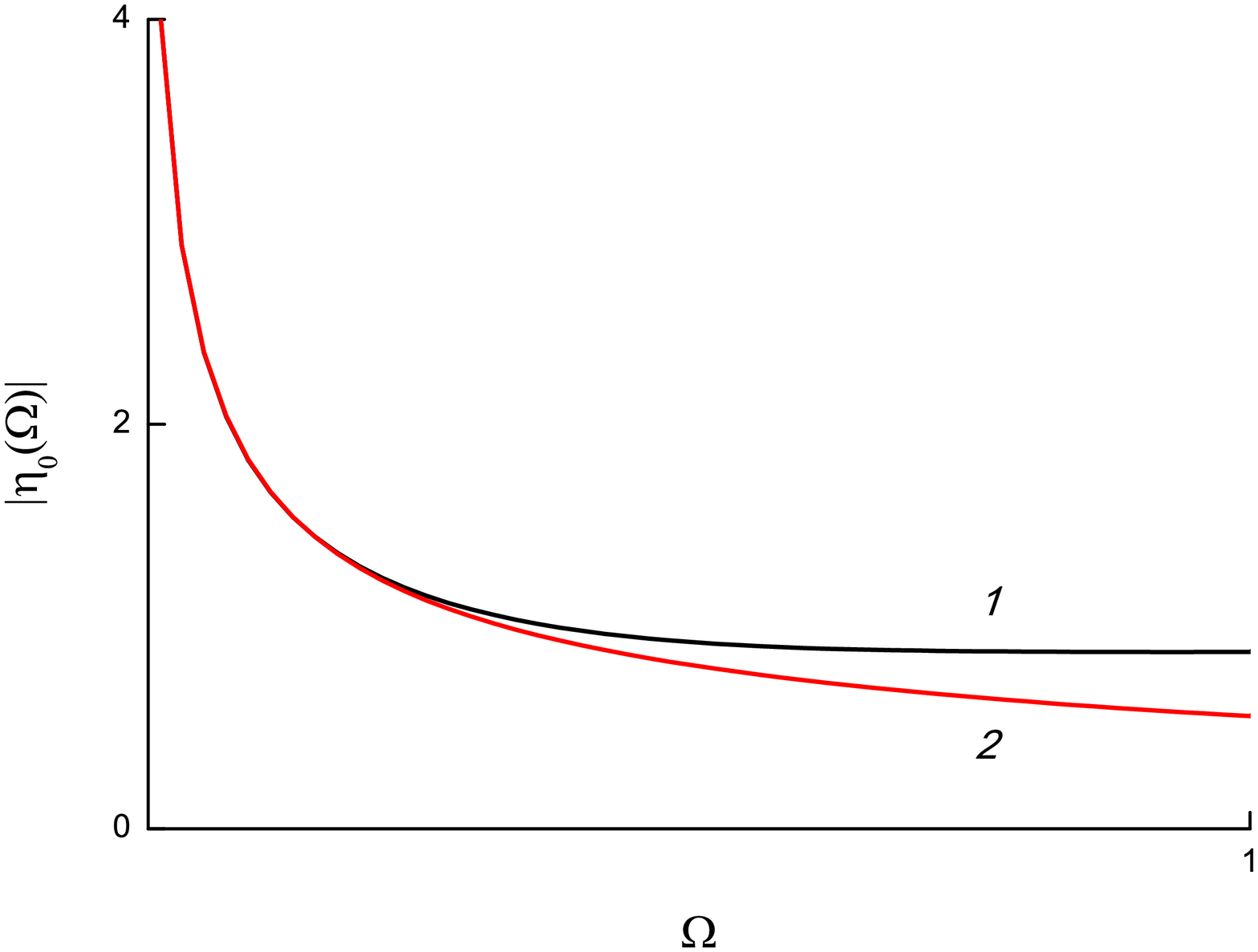}
\end{center}
\begin{center}
{{ Fig. 6. Comparison of modules of zero $\eta_0=\theta_0(\Omega)$ (curve 1) and
$\eta=\eta_0^{as}(\Omega)$\\ (curve 2), $\Pr=2/3$.}}
\end{center}
\end{figure}

Research of properties of zero of dispersion function is carry out on
to the formula (4.8). On fig. 5 it is spent
comparison of modules of exact value of zero
$ \eta_0(\Omega)$ (curve 1) and asymptotic representations
zero $ \eta_0^{as}$ (curve 2).

Let's enter function of errors, which is the function of  relative deviation
asymptotic representations of the module of zero from the module of its
exact representation
$$\boxed{\textcolor{Blue}{
O(\Omega)=
\dfrac{|\eta_0(\Omega)|-|\eta_0^{as}(\Omega)|}{|\eta_0(\Omega)|}
\cdot 100\%.}}
$$

On fig. 7 the behaviour of function of errors is presented as function
of dimensionless frequencies of oscillations of the plane limiting
rarefied gas in Stokes' second problem. From fig. 7 it is visible,
that in an interval
0$ \leqslant \Omega\leqslant 0.2$ quantity of function of errors not
exceeds one percent. This fact allows
in applied questions to use asymptotic representation of  zero of
dispersion function.
\begin{figure}[h]
\begin{center}
\includegraphics[width=17.0cm, height=10cm]{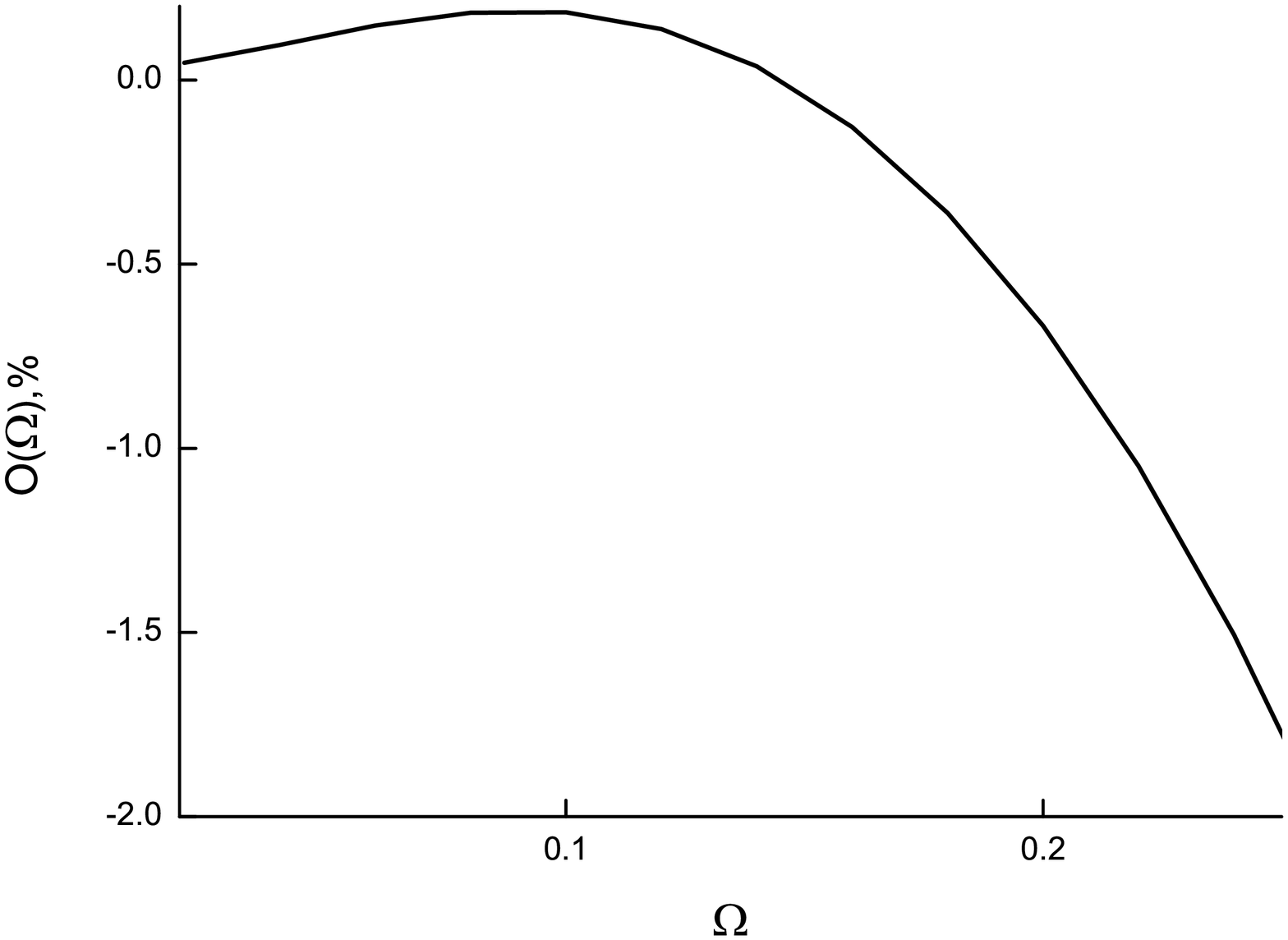}
\end{center}
\begin{center}
{{ Fig. 7. Relative deviations of the module of zero
$|\eta_0^{as}(\Omega)|$  from $|\eta_0(\Omega)|$, $\Pr=2/3$.}}
\end{center}
\end{figure}

\begin{center}
\bf 5. Conclusions
\end{center}
In the present work zero of dispersion function from the Stokes' second
problem are investigated. Stokes' second problem is the problem about
behaviour of the rarefied gas filling half-space. A plane, limitting
the half-space, makes harmonious oscillations in own plane.
It is used the linearization kinetic ellipsoidal statistical equation.
By means of the solution of boundary value Riemann problem the formula
of factorization of dispersion function is proved.
By means of the factorization formula in  explicit form
there are zero of dispersion function and their research of
dependence on quantity of dimensionless frequency of the plane limiting gas
is carried out.
The interval of values of frequency of oscillations of the plane
0$ \leqslant \Omega \leqslant 0.2$ in which the quantity of function
of errors does not exceed one percent is found out.

\end{document}